\documentclass[aps,preprint,amssymb]{revtex4}

\def\spa#1.#2{\langle#1\,#2\rangle}
\def\spb#1.#2{[#1\,#2]}
\def\spba#1.#2.#3{[#1|#2|#3\rangle}
\def\spab#1.#2.#3{\langle#1|#2|#3]}
\def\spaa#1.#2.#3.#4{\langle#1|#2#3|#4\rangle}

\usepackage{axodraw}

\newsavebox{\boxABCHsmall}
\savebox{\boxABCHsmall}(62.5,40){
\Boxc(0.000000,-62.000000)(15.000000,15.000000)
\Line(-7.500000,-69.500000)(-18.106602,-80.106602)
\put(-25.026019,-87.026019){\makebox(5.000000,5.000000){\tiny $i$}}
\Line(-7.500000,-54.500000)(-18.106602,-43.893398)
\put(-25.026019,-41.973981){\makebox(5.000000,5.000000){\tiny $i{+}1$}}
\Line(7.500000,-54.500000)(7.500000,-39.500000)
\put(5.000000,-35.750000){\makebox(5.000000,5.000000){\tiny $i{+}2$}}
\Line(7.500000,-54.500000)(13.240251,-40.641807)
\Line(7.500000,-54.500000)(18.106602,-43.893398)
\Line(7.500000,-54.500000)(21.358193,-48.759749)
\Line(7.500000,-54.500000)(22.500000,-54.500000)
\put(26.250000,-57.000000){\makebox(5.000000,5.000000){\tiny $j$}}
\Line(7.500000,-69.500000)(22.500000,-69.500000)
\put(26.250000,-72.000000){\makebox(5.000000,5.000000){\tiny $j{+}1$}}
\Line(7.500000,-69.500000)(21.358193,-75.240251)
\Line(7.500000,-69.500000)(18.106602,-80.106602)
\Line(7.500000,-69.500000)(13.240251,-83.358193)
\Line(7.500000,-69.500000)(7.500000,-84.500000)
\put(5.000000,-93.250000){\makebox(5.000000,5.000000){\tiny $i{+}n{-}1$}}
}

\newsavebox{\boxABCF}
\savebox{\boxABCF}(125,125){
\Boxc(0.000000,-62.000000)(30.000000,30.000000)
\DashLine(-5.000000,-62.000000)(-25.000000,-62.000000){3}
\DashLine(0.000000,-67.000000)(0.000000,-87.000000){3}
\DashLine(5.000000,-62.000000)(25.000000,-62.000000){3}
\DashLine(0.000000,-57.000000)(0.000000,-37.000000){3}
\Line(-15.000000,-77.000000)(-30.909903,-92.909903)
\put(-40.039029,-102.039029){\makebox(5.000000,5.000000){$1^-$}}
\Line(-15.000000,-47.000000)(-30.909903,-31.090097)
\put(-40.039029,-26.960971){\makebox(5.000000,5.000000){$2^-$}}
\Line(15.000000,-47.000000)(15.000000,-24.500000)
\put(12.500000,-17.625000){\makebox(5.000000,5.000000){$3^-$}}
\Line(15.000000,-47.000000)(30.909903,-31.090097)
\put(35.039029,-26.960971){\makebox(5.000000,5.000000){$4^-$}}
\Line(15.000000,-47.000000)(37.500000,-47.000000)
\put(44.375000,-49.500000){\makebox(5.000000,5.000000){$5^+$}}
\Line(15.000000,-77.000000)(37.500000,-77.000000)
\put(44.375000,-79.500000){\makebox(5.000000,5.000000){$6^+$}}
\Line(15.000000,-77.000000)(30.909903,-92.909903)
\put(35.039029,-102.039029){\makebox(5.000000,5.000000){$7^+$}}
\Line(15.000000,-77.000000)(15.000000,-99.500000)
\put(12.500000,-111.375000){\makebox(5.000000,5.000000){$8^+$}}
\put(16,-60){\makebox(10,10){$+$}}
\put(16,-75){\makebox(10,10){$-$}}
\put(-27,-60){\makebox(10,10){$-$}}
\put(-27,-75){\makebox(10,10){$+$}}
\put(-12.5,-46){\makebox(10,10){$+$}}
\put(2.5,-46){\makebox(10,10){$-$}}
\put(-12.5,-88){\makebox(10,10){$+$}}
\put(2.5,-88){\makebox(10,10){$-$}}
}

\newsavebox{\boxABCFsmall}
\savebox{\boxABCFsmall}(62.5,62.5){
\Boxc(0.000000,-31.250000)(15.000000,15.000000)
\Line(-7.500000,-38.750000)(-18.106602,-49.356602)
\put(-26.642136,-57.892136){\makebox(10.000000,10.000000){\tiny $1$}}
\Line(-7.500000,-23.750000)(-18.106602,-13.143398)
\put(-26.642136,-14.607864){\makebox(10.000000,10.000000){\tiny $2$}}
\Line(7.500000,-23.750000)(7.500000,-8.750000)
\put(2.500000,-8.750000){\makebox(10.000000,10.000000){\tiny $3$}}
\Line(7.500000,-23.750000)(18.106602,-13.143398)
\put(16.642136,-14.607864){\makebox(10.000000,10.000000){\tiny $4$}}
\Line(7.500000,-23.750000)(22.500000,-23.750000)
\put(22.500000,-28.750000){\makebox(10.000000,10.000000){\tiny $5$}}
\Line(7.500000,-38.750000)(22.500000,-38.750000)
\put(22.500000,-43.750000){\makebox(10.000000,10.000000){\tiny $6$}}
\Line(7.500000,-38.750000)(18.106602,-49.356602)
\put(16.642136,-57.892136){\makebox(10.000000,10.000000){\tiny $7$}}
\Line(7.500000,-38.750000)(7.500000,-53.750000)
\put(2.500000,-63.750000){\makebox(10.000000,10.000000){\tiny $8$}}
}

\newsavebox{\boxABCG}
\savebox{\boxABCG}(125,125){
\Boxc(0.000000,-62.000000)(30.000000,30.000000)
\DashLine(-5.000000,-62.000000)(-25.000000,-62.000000){3}
\DashLine(0.000000,-67.000000)(0.000000,-87.000000){3}
\DashLine(5.000000,-62.000000)(25.000000,-62.000000){3}
\DashLine(0.000000,-57.000000)(0.000000,-37.000000){3}
\Line(-15.000000,-77.000000)(-30.909903,-92.909903)
\put(-40.039029,-102.039029){\makebox(5.000000,5.000000){$1^-$}}
\Line(-15.000000,-47.000000)(-30.909903,-31.090097)
\put(-40.039029,-26.960971){\makebox(5.000000,5.000000){$2^-$}}
\Line(15.000000,-47.000000)(15.000000,-24.500000)
\put(12.500000,-17.625000){\makebox(5.000000,5.000000){$3^-$}}
\Line(15.000000,-47.000000)(26.250000,-27.514428)
\put(28.437500,-21.895440){\makebox(5.000000,5.000000){$4^-$}}
\Line(15.000000,-47.000000)(34.485572,-35.750000)
\put(40.104560,-33.562500){\makebox(5.000000,5.000000){$5^+$}}
\Line(15.000000,-47.000000)(37.500000,-47.000000)
\put(44.375000,-49.500000){\makebox(5.000000,5.000000){$6^+$}}
\Line(15.000000,-77.000000)(37.500000,-77.000000)
\put(44.375000,-79.500000){\makebox(5.000000,5.000000){$7^+$}}
\Line(15.000000,-77.000000)(15.000000,-99.500000)
\put(12.500000,-111.375000){\makebox(5.000000,5.000000){$8^+$}}
\put(16,-60){\makebox(10,10){$+$}}
\put(16,-75){\makebox(10,10){$-$}}
\put(-27,-60){\makebox(10,10){$-$}}
\put(-27,-75){\makebox(10,10){$+$}}
\put(-12.5,-46){\makebox(10,10){$+$}}
\put(2.5,-46){\makebox(10,10){$-$}}
\put(-12.5,-88){\makebox(10,10){$+$}}
\put(2.5,-88){\makebox(10,10){$-$}}
}

\begin{document}

\hfill hep-th/0412265

\hfill NSF--KITP--04-135

\hfill PUPT-2147

\bigskip
\bigskip

\title{Dissolving ${\mathcal{N}} = 4$ loop amplitudes into QCD tree amplitudes}

\bigskip

\bigskip
\author{Radu Roiban}
\affiliation{
Department of Physics, Princeton University\\
Princeton, NJ 08544, USA\\
{\tt rroiban@princeton.edu}
}

\author{Marcus Spradlin and Anastasia Volovich}
\affiliation{ 
Kavli Institute for Theoretical Physics\\
Santa Barbara, CA 93106, USA\\
{\tt spradlin, nastja@kitp.ucsb.edu}
}

\bigskip
\bigskip

\begin{abstract}
We use the infrared consistency of one-loop 
amplitudes in ${\mathcal{N}} = 4$ Yang-Mills theory
to derive a compact analytic formula for a
tree-level NNMHV gluon scattering amplitude in QCD,
the first such formula known.
We argue that the infrared  conditions, coupled with
recent advances in calculating one-loop
box coefficients, can 
give a new tool for computing tree-level amplitudes in general.
Our calculation suggests that many
amplitudes have a structure which is even simpler than that 
revealed so far by current twistor-space constructions.
\end{abstract}

\maketitle

\section{Introduction}

Gluon scattering amplitudes in Yang-Mills theory
have many remarkable properties.
Witten discovered that they are localized on curves
in twistor space \cite{Witten:2003nn}
and proposed an interpretation
for this fact
in terms of a twistor string theory.
There has been great success in calculating tree-level
\cite{Witten:2003nn,Roiban:2004vt,Roiban:2004ka,Roiban:2004yf,Cachazo:2004kj,Gukov:2004ei,Bena:2004ry} 
amplitudes
using twistor-inspired technology, including the construction
of a one-line (though implicit) formula for the complete tree-level
S-matrix 
\cite{Roiban:2004yf}.

Twistor string theory led in \cite{Cachazo:2004kj} to the
development of a set of diagrammatic 
rules for the calculation of tree-level amplitudes which are
far simpler than standard  Feynman diagrams.
Nevertheless, there is reason to believe that the formulas
generated by the
CSW rules
grow in complexity much more quickly than the
underlying amplitudes do.  In this paper we report
a compact analytic representation for the simplest
NNMHV amplitude
\footnote{Our conventions are mostly standard, except that we
use the `twistor' convention for the negative helicity inner product,
which differs from the convention of \cite{Bern:2004ky} by a sign.
In particular, we use
$\spa{p}.{q} \spb{p}.{q} = 2 p \cdot q$,
$\spba{p}.{q}.{r} = \spb{p}.{q} \spa{q}.{r}$,
$\spba{p}.{(q+r)}.{s} = \spba{p}.{q}.{s} +
\spba{p}.{r}.{s}$,
etc., and define
$k_i^{[r]} = p_i + \cdots p_{i+r-1}$
and $t_i^{[r]} = (k_i^{[r]})^2$.}
\begin{eqnarray}
\label{eq:mainresult}
&&A(1^-,2^-,3^-,4^-,5^+,6^+,7^+,8^+) =
 \frac{{\spba5.{k_2^{[3]}}.1}{}^3}
{t_2^{[4]}
\spb2.3 \spb3.4 \spb4.5 \spa6.7 \spa7.8 \spa8.1 \spba2.{k_3^{[3]}}.6}
\cr
&& -
\frac{1}{t_7^{[3]}\spa7.8 \spa8.1 \spba3.{k_4^{[2]}}.6
\spab7.{k_8^{[2]}}.2}
\left[
\frac{\langle 1|k_7^{[2]}k_3^{[4]}k_3^{[2]}|5]^3}
{t_3^{[3]} t_3^{[4]} \spb3.4 \spb4.5
\spba2.{k_3^{[3]}}.6}
+ \frac{{\spaa1.{k_7^{[2]}}.{k_5^{[2]}}.4}{}^3}
{t_4^{[3]} \spb2.3
\spa4.5 \spa5.6}
\right]  + \overline{\rm flip},
\end{eqnarray}
which otherwise would require summing 44 CSW diagrams to write down
\footnote{Some savings could be achieved by writing down the amplitude
from an `intermediate prescription,' as
outlined in \cite{Bena:2004ry}.
Given the miraculous cancellations we observe
in this paper, it is natural to wonder whether there is a similar
cancellation between the terms appearing
in such an expression, or between some of the 44
CSW diagrams.  However, all of these terms individually depend
on an arbitrary reference spinor $\eta$ (which drops out of the full
amplitude), and while it is likely that cancellations occur for certain
clever choices of $\eta$, is it not a priori clear how to choose
$\eta$ to maximize the cancellations.}.
Previously, no such compact expression was known for any NNMHV amplitude.
The $``+ {\overline{\rm flip}}$'' denotes that this full expression
should be added to its image under
the operation which relabels $i \to 9 - i$ and simultaneously
exchanges $\langle\,\rangle$ and $[\,]$.

Gluon scattering amplitudes are
obviously of phenomenological interest as they are the
basic building block
for computing QCD backgrounds to jet production.  From
a theoretical standpoint,
perhaps the most
interesting fact about the formula~(\ref{eq:mainresult})
is that  such a compact representation of this amplitude
exists at all.
The existence of this formula
suggests that there is 
hidden structure and simplicity underlying
tree amplitudes
even beyond what the CSW rules expose
\footnote{This suggestion has also 
been made in \cite{hepth:0412210}, which appeared
as this article was in preparation.}.
Moreover, the formula~(\ref{eq:mainresult}) is much simpler
than we had any right to expect, given the 
procedure used to compute it.  Rather, (\ref{eq:mainresult}) hinges
upon the miraculous cancellation of a large number of terms,
which at the moment we cannot explain.
The lesson we would like to draw from this exercise is that
there still is a lot to learn about the structure of Yang-Mills amplitudes,
even at tree-level, and in particular
that CSW is not the end of the story \footnote{It would be interesting
to see how many roots there are to the algebraic equations
which describe this amplitude as explained in \cite{Roiban:2004yf}.
Perhaps there is some relation between the number of roots and the number
of terms in the most compact expression possible for a given amplitude.}.

Recently there has also been much progress on the computation of
one-loop amplitudes 
\cite{Bern:1994cg,Bern:1994zx,Britto:2004nc,Bern:2004ky,
Bena:2004xu,Cachazo:2004by,Cachazo:2004dr,
Brandhuber:2004yw,Luo:2004ss,Luo:2004nw,Bidder:2004tx,
Dixon:2004za,Quigley:2004pw,Bedford:2004nh,Britto:2004nj,hepth:0412210}.
In particular, a prescription has recently been 
developed \cite{Britto:2004nc}
for calculating the coefficient
of any one-loop box function in the ${\mathcal{N}} = 4$ theory,
based on the notion of generalized cuts
\cite{Bern:1997sc,Bern:2000dn,Bern:2004ky,Bern:2004cz}
(in this case, quadruple cuts).
One-loop amplitudes have infrared (IR) singularities, and the leading
singularities are proportional to tree amplitudes in a way
that we review below. Enforcing this fact may provide, as pointed out 
in \cite{Bern:2004ky}, new and more compact expressions for tree-level
scattering amplitudes.
We derived the tree-level formula~(\ref{eq:mainresult}) by
computing several box coefficients appearing in the
one-loop amplitude corresponding to~(\ref{eq:mainresult})
and then reading 
off its IR singularity.
In the next section we
review the necessary ingredients
and discuss a couple of strategies for maximizing
the efficiency of this kind of computation.

\section{Trees from Loops}

The set of dimensionally regularized
integrals which can appear at one-loop
has been completely classified.
Any $n$-gluon partial amplitude in
$(4 - 2\epsilon)$-dimensional ${\cal N} = 4$ SYM theory can be
expressed in terms of
${n \choose 4}$  box functions
$B_n(i,j,k,l)$:
\begin{equation}
\label{eq:generalamplitude}
A_{n;1\,{\rm loop}} = i c_\Gamma (\mu^2)^\epsilon
\sum c_{n;ijkl} B_n(i,j,k,l), \qquad
c_\Gamma = \frac{1}{(4 \pi)^{2 - \epsilon}}
\frac{\Gamma(1 + \epsilon) \Gamma^2(1 - \epsilon)}
{\Gamma(1 - 2 \epsilon)},
\end{equation}
where $\mu$ is a renormalization scale.
The box functions are completely known (but complicated)
functions of the momenta $p_i$ of the $n$ gluons.
Therefore the decomposition~(\ref{eq:generalamplitude})
reduces the problem of calculating any one-loop amplitude to the problem
of calculating the (relatively much simpler)
box coefficients $c_{n;ijkl}$, which can depend
on the momenta as well as the helicities of the $n$ gluons.

The box functions
are typically transcendental functions
of the momenta (involving logarithms
and dilogarithms).
They are most conveniently labeled \cite{hepth:0412210} by
a set of distinct, ordered indices $i,j,k,l$ chosen from the set
$\{1,\ldots,n\}$.
For example,
\begin{eqnarray}
\label{eq:examplebox}
B_8(1,2,3,6) &=&
{\hbox{\lower 30pt\hbox{
\begin{picture}(50,60)(10,0)
\put(0,0){\usebox{\boxABCFsmall}}
\end{picture}
}}}
=
- \frac{1}{2 \epsilon^2}
\left[ (-t_1^{[2]})^{-\epsilon}
+ 2 (-t_2^{[4]})^{-\epsilon} -
(-t_3^{[3]})^{-\epsilon}
- (-t_6^{[3]})^{-\epsilon}\right]
\cr
&&
+~{\rm Li}_2\left(1 - \frac{t_3^{[3]}}{t_2^{[4]}}\right)
+ {\rm Li}_2\left(1 - \frac{t_6^{[3]}}{t_2^{[4]}}\right)
+ \frac{1}{2} \ln^2 \left( \frac{t_1^{[2]}}{t_2^{[4]}}\right) +
{\cal{O}}(\epsilon).
\end{eqnarray}
The indices indicate the four external legs which immediately
follow
(in a clockwise sense, conventionally)
the four internal propagators in the corresponding box integral.
Explicit formulas for all box functions can be found in any standard
reference, and we will not repeat them all here but rather
provide~(\ref{eq:examplebox})  as an illustrative example for
what follows.

Each box function contains ${\cal{O}}(1/\epsilon)$ terms of the
form $-\ln(-t_i^{[r]})/\epsilon$ for various $i$ and $r$, with
coefficients $0$, $\pm 1$ or $\pm \frac{1}{2}$.
Meanwhile,
the ${\cal{O}}(1/\epsilon)$ IR singularity in any one-loop amplitude
is known on general grounds to be
\cite{Giele:1991vf,Kunszt:1994mc,Catani:1998bh}
\begin{equation}
\label{eq:ir}
A_{n;1\,{\rm loop}}|_{\epsilon~{\rm pole}} = - \frac{c_\Gamma}{\epsilon^2}
\sum_{i=1}^n \left(\frac{\mu^2}{-t_i^{[2]}}\right)^\epsilon A_n^{\rm tree}.
\end{equation}
One constraint on the box coefficients is that they conspire
in such a way that the IR divergences in the box functions combine
together so that this equation is satisfied.
A set of $n(n-3)/2$ linear equations, $n$ of which involve
the tree amplitude, can be deduced by equating~(\ref{eq:generalamplitude})
to~(\ref{eq:ir}) and
reading off the coefficients of the various $-\ln(-t_i^{[r]})/\epsilon$
poles.

It is likely that in many cases, the IR equations implied
by~(\ref{eq:ir}) allow one to write down explicit formulas
for tree amplitudes where none was previously known, or to
write down more compact formulas for previously known
amplitudes.
An example of the latter is the remarkable formula
\begin{equation}
A(1^-,2^-,3^-,4^+,5^+,6^+)
= \frac{{\spab1.{k_2^{[2]}}.4}{}^3}{t_2^{[3]}
\spb2.3 \spb3.4 \spa5.6 \spa6.1 [2|k_3^{[2]}|5\rangle} + 
\frac{{[6|k_1^{[2]}|3\rangle}{}^3}{t_6^{[3]} \spb2.1 \spb1.6 \spa5.4
\spa4.3  [2|k_6^{[2]}|5\rangle},
\end{equation}
which follows from the collinear limit of a three-term
representation of the seven particle
amplitude $A(1^-,2^-,3^-,4^+,5^+,6^+,7^+)$
found in \cite{Bern:2004ky} using the IR equations for $n=7$.

A particularly compact formula
which follows from~(\ref{eq:ir})
is
\begin{equation}
\label{eq:simpleir}
A^{\rm tree}(1,\ldots,n) = \frac{1}{2}
\sum_{j=i+2}^{i+n-2}
{\rm coefficient}
\left[
{\hbox{\raise 0pt\hbox{
\begin{picture}(60,40)(0,0)
\put(-5,0){\usebox{\boxABCHsmall}}
\end{picture}
}}}
\right],
\end{equation}
for any $i$.
This formula expresses an arbitrary tree-level amplitude as
a sum of $n-3$ box coefficients of the corresponding one-loop
amplitude.
For $n=8$ we have verified that no linear combination
of the IR equations
allows one to extract the tree amplitude from fewer than
$n-3=5$
coefficients.
We conjecture that this remains true for any $n$, so that~(\ref{eq:simpleir})
is in a sense the most efficient linear combination of IR equations
possible.
Of course, not all box coefficients are equally 
simple to compute, so in practice it may be worthwhile to consider
a linear combination of IR equations which has a larger number of
terms which are however
individually simpler to compute.
We will encounter an example of this in the next
section.

\section{The $A(1^-,2^-,3^-,4^-,5^+,6^+,7^+,8^+)$ Amplitude}

In this section we investigate the tree-level eight-gluon amplitude
\begin{equation}
A_8 \equiv A(1^-,2^-,3^-,4^-,5^+,6^+,7^+,8^+),
\end{equation}
for which no compact analytic formula was previously known.
As a preliminary remark we note that $A_8$ possesses two commuting
${\mathbb{Z}}_2$
symmetries.  One is
the symmetry under the flip operation defined by
\begin{equation}
F[X] = X(1 \leftrightarrow 4, 2 \leftrightarrow 3,
5 \leftrightarrow 8,  6 \leftrightarrow 7).
\end{equation}
The other is a conjugation symmetry combined with a relabeling
of the indices, which we will denote by
\begin{equation}
G[X] = \overline{X}(1 \leftrightarrow 5, 2 \leftrightarrow 6,
3 \leftrightarrow 7, 4 \leftrightarrow 8).
\end{equation}
The bar over the $X$ denotes the reversal of the helicities
of the inner products, i.e.~$\langle\,\rangle \leftrightarrow [\,]$.

In order to calculate $A_8$ we use
the IR equations introduced in the
previous section, combined with the
new technology of \cite{Britto:2004nc}
for calculating one-loop box coefficients using quadruple cuts.
In this approach, the coefficient of a box function is given
by the product of the four tree amplitudes sitting at the corners
of the box.  
This construction implies that
the complexity of a box coefficient is indicated by the complexity
of the tree amplitudes sitting at the corners of the box.
Moreover, when coupled with the IR equations,
it implies that various tree amplitudes satisfy 
recursion relations relating $n$-particle amplitudes
to amplitudes with fewer particles.
In general these relations are quartic, although we see
that the choice~(\ref{eq:simpleir}) actually renders them
quadratic since only two of the four corners involve a non-trivial
lower-point tree amplitude.

In particular,
the simplest box coefficients are those which
have an MHV (or $\overline{\rm MHV}$) amplitude on each corner,
which is guaranteed to happen
when each corner has fewer than four external
legs.  Let us refer to  these as ``$({\rm MHV})^4$ boxes''.
For $n = 8$, it turns out that there is an almost unique 
\footnote{There is one
other linearly independent equation with only $({\rm MHV})^4$ boxes:
$c_{1245} - c_{1258} - c_{1456} + c_{1568} = 0$.
However, 
the holomorphic anomaly in the $t_1^{[4]}$ cut 
can be used to
verify that all four of these coefficients are equal to each other (and
nonzero), so this equation is not useful.}
linear combination of IR equations which expresses
the desired tree amplitude $A_8$ 
in terms of the coefficients of $({\rm MHV})^4$ boxes only:
\begin{equation}
\label{eq:coolone}
4 A_8 = X + F[X] + G[X] + F[G[X]],
\end{equation}
where
\begin{equation}
X = X_1 =
c_{1236} + c_{1346} + c_{1347}
+ \frac{1}{2} c_{2347} + c_{2367} + c_{2368}.
\end{equation}
Although it is intriguing that it is possible to
express the tree amplitude $A_8$ in a manifestly symmetric
manner using only $({\rm MHV})^4$ boxes, it is not
clear that~(\ref{eq:coolone}) would provide the most
compact representation of $A_8$.

An alternate representation of $A_8$ is given by the same
formula~(\ref{eq:coolone}), but with
\begin{equation}
\label{eq:otherx}
X = X_2 = c_{1236} + c_{1237} + c_{1246} + c_{1247}
+ \frac{1}{2} c_{1258},
\end{equation}
where we omitted a term $c_{1256}$ which is easily
seen to be zero by analyzing the holomorphic anomaly of
the $t_1^{[4]}$ cut.  We have worked out all of the coefficients
appearing in~(\ref{eq:otherx}), and the purpose
of this note is to
point out a
most unexpected surprise:  almost all of the terms appearing
on the right-hand side of~(\ref{eq:otherx}) cancel (after summing
over the ${\mathbb{Z}}_2 \times
{\mathbb{Z}}_2$ images in~(\ref{eq:coolone})).
The only terms remaining are
\begin{equation}
\label{eq:xtwo}
X_2 = - \frac{2}{t_1^{[2]} t_2^{[4]}} b_{1236}^{+-}
- \frac{2}{t_1^{[2]} t_7^{[3]}} b_{1237}^{+-},
\end{equation}
where the $b$'s are 
partial contributions to the 1236 and 1237 integral coefficients.
In particular, they denote the following quadruple cuts:
\begin{equation}
b_{1236}^{+-} =
{\hbox{\raise -55pt\hbox{
\begin{picture}(100,125)(20,0)
\put(0,0){\usebox{\boxABCF}}
\end{picture}
}}},\qquad
b_{1237}^{+-} =
{\hbox{\raise -55pt\hbox{
\begin{picture}(100,125)(20,0)
\put(0,0){\usebox{\boxABCG}}
\end{picture}
}}},
\end{equation}
which we computed using the methods of \cite{Britto:2004nc}, giving
\begin{eqnarray}
b^{+-}_{1236}&=& \frac{t_1^{[2]} {\spba5.{k_6^{[3]}}.1}{}^3}
{\spb2.3 \spb3.4 \spb4.5 \spa6.7 \spa7.8 \spa8.1 \spba2.{k_3^{[3]}}.6},\cr
b^{+-}_{1237}&=&  \frac{t_1^{[2]}}{\spa7.8 \spa8.1 \spba3.{k_4^{[2]}}.6
\spab7.{k_8^{[2]}}.2}
\left[
\frac{\langle 1|k_7^{[2]}k_3^{[4]}k_3^{[2]}|5]^3}
{t_3^{[3]} t_3^{[4]} \spb3.4 \spb4.5
\spba2.{k_3^{[3]}}.6}
+ \frac{{\spaa1.{k_7^{[2]}}.{k_5^{[2]}}.4}{}^3}
{t_4^{[3]} \spb2.3
\spa4.5 \spa5.6}
\right].
\end{eqnarray}
The 1236 and 1237 boxes each receive a contribution from a second
helicity assignment which we have not drawn here and which is not
needed in~(\ref{eq:xtwo}).  The other factors
appearing in~(\ref{eq:xtwo}) are the usual conversion
factors between integral coefficients and box coefficients.

In fact, we have verified that $X_2$ satisfies the non-manifest relation
\begin{equation}
\label{eq:symmetry}
X_2 + F[G[X_2]] = F[X_2] + G[X_2]
\end{equation}
which renders two of the terms in~(\ref{eq:coolone}) redundant,
as anticipated in~(\ref{eq:mainresult}) (where we use
$\overline{\rm flip}$ to denote the composition $F[G[~]]$).
In conclusion, let us note that
we have verified numerically that 
the formula~(\ref{eq:mainresult})
agrees with the amplitude
obtained by summing up the necessary 44 CSW diagrams.
As further evidence, it is rather straightforward to check
that~(\ref{eq:mainresult}) has all of the correct collinear limits.
In particular, it reproduces the simple three-term
representation of the
7-particle tree amplitude
found in \cite{Bern:2004ky}, as one might expect from taking the
collinear limits of the full box coefficients entering
into the calculation.

\begin{acknowledgments}
We would like to thank Zvi Bern for many helpful
discussion and correspondence, and especially for
sending us the list of IR equations for $n=8$.
R.~Roiban is grateful to the KITP for hospitality and support
during the QCD and String Theory Workshop, where
this work was initiated.
This research was supported in part by the National
Science Foundation under
Grant No.~PHY99-07949 (MS, AV), and by the DOE
under Grant No.~DE-FG02-91ER40671 (RR).
\end{acknowledgments}

\end{document}